\documentclass[pra,twocolumn,superscriptaddress,showpacs]{revtex4}
\usepackage{amsmath}
\usepackage{amsfonts}
\usepackage{graphicx}

\begin{document}

\title{Multipartite entangled coherent states}
\author{Xiaoguang Wang}
\affiliation{Institute of Physics and Astronomy, University of Aarhus,
	DK-8000, Denmark}
\affiliation{Institute for Scientific Interchange (ISI) Foundation,
 Viale Settimio Severo 65, I-10133 Torino, Italy}
\author{Barry C.\ Sanders}
\affiliation{Department of Physics, Macquarie University,
Sydney, New South Wales 2109, Australia}
\date{\today}

\begin{abstract}
We propose a scheme for generating multipartite entangled coherent
states via entanglement swapping, with an example of a physical
realization in ion traps.
Bipartite entanglement of these multipartite states is
quantified by the concurrence. We also use the $N$--tangle 
to compute multipartite entanglement for certain
systems. Finally we establish that these results for entanglement
can be applied to more general multipartite entangled nonorthogonal states.
\end{abstract}

\pacs{03.65.Ud, 03.67. Hk, 03.67.Lx}
\maketitle

\section{Introduction}

Quantum entanglement is at the heart of quantum information theory and plays
a key role in quantum information such as quantum teleportation \cite{Ben93}, 
superdense coding \cite{Ben92},
quantum key distribution \cite{Eke91},
and telecoloning \cite{Mur99}. Genuine entanglement arises when the state of
a multipartite system is nonseparable. Despite extensive efforts to quantify
entanglement \cite{Wer89,Hil97}, characterization and
classification of general mixed entangled states remains
an important
challenge. Entangled nonorthogonal states are even more challenging than
their well-studied orthogonal counterparts, yet have not received
the same degree of attention despite the importance of
nonorthogonality to quantum theory and the importance of 
entanglement of nonorthogonal states to quantum information 
such as quantum key distribution \cite{Fuc97}.
The entangled coherent state
(ECS) \cite{San92,Wie93,Man95,San95,San00},
or multipartite superposition of coherent states \cite{Cha92,Ans94},
is the most
well--known example of entangled nonorthogonal states, along with
the related entangled
squeezed states \cite{San95} and entangled SU(2) and SU(1,1) coherent
states \cite{Wan00}.

The bipartite ECS can exhibit various nonclassical properties such as
sub--Poissonian
statistics, two-mode squeezing and violations of the Cauchy--Schwarz
inequalities \cite{Cha92},
as well as violating Bell's inequality \cite{San92,Man95}.
Although most attention has been devoted to the bipartite ECS,
the multipartite case has been studied as well \cite{Ans94},
but not to the extent we undertake here, 
namely generating such states, discussing their potential realization
in ion traps and quantifying the degree of entanglement.
Such studies of the ECS are of interest beyond obtaining a fundamental
understanding of nonorthogonal entangled states: such states can 
be employed in quantum information theory, and in quantum
computing applications in particular \cite{Mun00}.
This particular application is relevant to our discussion of the multipartite 
entangled coherent state (MECS), as it considers qubits encoded
as two--dimensional center--of--mass (CoM) vibrational motion of
two ions in an ion trap.  These qubits are measured by swapping entanglement
from the vibrational to the internal states of the ion \cite{Mun00}.
Our analysis of the MECS could enable generalizations of such qubits to
large entangled systems.

Here we discuss the MECS and propose a scheme to generate
such states via entanglement swapping, and we calculate the
degree of entanglement by employing concurrence \cite{Hil97}.
We also compute the $N$-tangle \cite{Cof00,Won01} to
characterize the multipartite entanglement of the tripartite and 
even--number MECS.

\section{Generation of multipartite entangled coherent states}

We consider a set of $N$ ions in a linear trap
coupled together via the Coulomb interaction.
The collective motion of the ions is described by the dynamics
of the normal modes, with~$a_i$ ($a_i^\dagger$) the annihilation
(creation) operator for the
$i^{\rm th}$ mode and $\nu_i$ the angular frequency of the $i^{\rm th}$
mode, $i \in \{1,2,\ldots,N\}$.
The fundamental mode ($i=1$) is the CoM mode with frequency
$\nu_1$, and the $i=2$ mode is the breathing mode with angular
frequency $\nu_2=\sqrt{3}\nu_1$.  Frequencies for
higher-order normal modes can be calculated~\cite{Jam98}.
We propose a method for generating entangled coherent states
of the normal phonon modes for the trapped ions.

The ions behave as effective two-level systems, with
$\vert 0 \rangle$ and $\vert 1 \rangle$ being the lower and upper
states. Unitary transformations that create superpositions of these
two states are generated by the Pauli raising and lowering operators
$\sigma_\pm$ and the inversion operator~$\sigma_z$.
In $^9$Be$^+$, the states $\vert 0 \rangle$ and $\vert 1 \rangle$
exist as two hyperfine sublevels with superpositions created via
stimulated Raman transitions~\cite{Kie01}.

The driving frequency between states $\vert 0 \rangle$
and $\vert 1 \rangle$ can be modified to excite one of the normal
modes of oscillation.  The normal mode can be excited by driving any 
one of the atoms, and, for a given atom $j$, an effective interaction
Hamiltonian between the $i^{\rm th}$ phonon mode and the $j^{\rm th}$ ion
is~\cite{Mon96,Ger97}
\begin{equation}
H_{ij}=g_{ij} a_i^{\dagger }a_i \sigma_{jz}~.
\label{interaction}
\end{equation}
where $g_{ij}$ is the coupling constant.
By choosing the duration, strength and frequency of the Raman pulse 
appropriately, the coupling strength can
be set to the same magnitude. Thus we can consider a fixed coupling strength
$g$. This means that any ion can be coupled to any of the normal modes of
oscillation, with the choice of mode and coupling strength determined
by experimental parameters.
Thus, $g_{ij}=g$ in Eq.~(\ref{interaction}).
Although challenging, this coupling between
ions and normal mode oscillations is possible.

A judicious choice of initial state will lead, by Hamiltonian
evolution, to the MECS.  We begin with the coherent state of the
vibrational collective mode.
If the phonon mode is initially in the ground state, 
the driving field is first configured to impart a displacement to the
chosen normal mode.
We can therefore assume that the $i^{\rm th}$ normal mode has been
prepared in a coherent state~$\vert\alpha\rangle_i$, and the ion $i$ can
be prepared in a superposition of $\vert 0 \rangle_i$ and $\vert 1 \rangle_i$.
The initial state for the $i^{\rm th}$ ion and the $i^{\rm th}$ normal mode is
prepared as
$\vert\psi (0)\rangle_i
	=2^{-1/2}\vert\alpha \rangle_i
	\otimes (\vert 0 \rangle_i+\vert 1 \rangle_i)$.
This state undergoes the Hamiltonian evolution $H_{ii}$~(\ref{interaction}) to yield
the time--dependent state (in the interaction picture) 
\begin{equation}
\vert\psi (\tau)\rangle_i=2^{-1/2}\left(\vert\alpha
	e^{i\tau}\rangle_i\otimes
	\vert 0\rangle_i+\vert\alpha e^{-i\tau}\rangle_i
	\otimes \vert 1\rangle_i\right) 
\label{tau}
\end{equation}
with $\tau=g t$ a normalized unit of time, scaled to the coupling strength~$g$.

State~(\ref{tau}) involves just one degree of freedom for the
vibrational mode. We can consider two identical ions, ions~1 and~2,
each prepared in the superposition state $\vert 0 \rangle$ and $\vert 1
\rangle$.  Let us also assume that the CoM mode 
and the breating mode have been prepared in identical
coherent states, that is, with the same amplitude and phase; the
two-mode coherent state for these two normal modes oscillation is
$\vert\alpha\rangle_1\otimes\vert\alpha\rangle_2$.  We then couple
ion~1 to the fundamental mode and ion~2 to the breathing mode by
adjusting the appropriate parameters of the two Raman beams, with
beam~1 directed at ion~1 and beam~2 directed at ion~2.  Moreover, the
coupling strength~$g$ between ion~1 and the CoM mode is
equal to the coupling strength between ion~2 and the breathing mode,
with both interactions occurring simultaneously; i.e., the interaction
Hamiltonian is
$H=g(a_1^\dagger a_1\sigma_{1z}+a_2^\dagger a_2\sigma_{2z}).$ Then the
Hamiltonian~(\ref{interaction}) applies to the coupling between ion~1
and normal mode~1 and the coupling between ion~2 and normal mode~2
simultaneously.  After a time~$\tau=gt$, the product state
\begin{multline}
\label{factorizable}
\vert\psi(\tau)\rangle_1\otimes\vert\psi(\tau)\rangle_2	\\
=2^{-1}(\vert\alpha
e^{i\tau}\rangle_1\otimes \vert\alpha e^{i\tau }\rangle_2\otimes\vert 0\rangle
_1\otimes\vert 0\rangle_2  \\
+\vert\alpha e^{-i\tau}\rangle_1\otimes |\alpha e^{-i\tau }\rangle_2\otimes
\vert 1\rangle_1\otimes |1\rangle_2  \\
+\vert\alpha e^{i\tau}\rangle_1\otimes |\alpha e^{-i\tau }\rangle_2\otimes
\vert 0\rangle_1\otimes\vert 1\rangle_2  \\
+\vert\alpha e^{-i\tau}\rangle_1\otimes |\alpha e^{i\tau }\rangle_2\otimes
\vert 1\rangle_1\otimes\vert 0\rangle_2)
\end{multline}
eventuates and is factorizable because two independent states have each
undergone independent evolutions.

The factorizable state~(\ref{factorizable}) may be transformed
to an entangled state via a suitable measurement process on the electronic
states of the two ions.
Direct measurements of the electronic states will not suffice to
entangle the vibrational states, but Bell state measurements of the
electronic states will work. This technique of Bell state measurements
to entangle other states is the hallmark of entanglement
swapping~~\cite{Zuk93}, and we apply this method to entanglement
swapping for nonorthogonal states.

Measurements must be performed with respect to the Bell state bases for
joint electronic states of both ions. The Bell states are defined to be 
\begin{mathletters}
\begin{eqnarray}
\vert\Phi ^{\pm }\rangle_{12} &=&2^{-1/2}\left(\vert 0\rangle_1\otimes |0\rangle
_2\pm\vert 1\rangle_1\otimes |1\rangle_2\right) ,  \\
\vert\Psi ^{\pm }\rangle_{12} &=&2^{-1/2}\left(\vert 0\rangle_1\otimes \vert 1\rangle
_2\pm\vert 1\rangle_1\otimes \vert 0\rangle_2\right) ~.
\label{eq:bell}
\end{eqnarray}
These Bell state measurements on the state (\ref{factorizable})
yield the normalized ECSs
\end{mathletters}
\begin{mathletters}
\begin{equation}
\frac{\vert\alpha e^{i\tau }\rangle_1\otimes \vert\alpha e^{i\tau
}\rangle_2\pm \vert\alpha e^{-i\tau }\rangle_1\otimes |\alpha e^{-i\tau
}\rangle_2}{\sqrt{2\pm 2\exp (-4\vert\alpha |^2\sin ^2\tau )
\cos [2\vert\alpha |^2\sin (2\tau )]}}
\label{eq:ecs_a}
\end{equation}
and
\begin{equation}
\frac{\vert\alpha e^{i\tau }\rangle_1\otimes |\alpha e^{-i\tau
}\rangle_2\pm \vert\alpha e^{-i\tau }\rangle_1\otimes |\alpha e^{i\tau
}\rangle_2}{\sqrt{2\pm 2\exp (-4\vert\alpha |^2\sin ^2\tau )}}~,
\label{eq:ecs_b}
\end{equation}
respectively.  For the specific case that $\tau =\pi /2$, Eqs.~(\ref{eq:ecs_a})
and~(\ref{eq:ecs_b}) reduce to the even and odd ECS~\cite{Cha92}
\end{mathletters}
\begin{equation}
\frac{|i\alpha \rangle_1\otimes |i\alpha \rangle_2\pm\vert-i\alpha
	\rangle_1\otimes\vert-i\alpha\rangle_2}
	{\sqrt{2\pm 2\exp (-4\vert\alpha\vert^2)}}~.
\end{equation}

Thus far we have considered only the single--particle and bipartite cases.
As the purpose of this paper is to develop and study the MECS, beyond the
bipartite case, we wish to generalize the above analysis from $N=2$
particles, or ions, to arbitrary~$N$. We therefore consider $N$ systems,
each prepared in the identical state $\vert\psi (\tau )\rangle_i$, i.e., 
\begin{equation}
\vert\psi (\tau )\rangle_1\otimes \vert\psi (\tau )\rangle_2\otimes \cdots \otimes
\vert\psi (\tau )\rangle_N~.
\end{equation}
By analogy with the Bell state measurements, in the multiparticle case,
measurements are performed with respect to the maximally entangled
multipartite electronic states 
\begin{multline}
2^{-1/2}(|i_1\rangle_1\otimes |i_2\rangle_2\otimes \cdots \otimes
\vert i_N\rangle_N	\\
\pm |\overline{i_1}\rangle_1\otimes |\vert\overline{i_2}\rangle_2\otimes \cdots 
	\otimes \vert\overline{i_N}\rangle_N)~,  \label{eq:e}
\end{multline}
for $i_k\in \{0,1\}$, $k\in \{1,2,\ldots,N\}$ and
$\overline{i_k} \equiv 1-i_k$.
The result of this measurement on the above state collapses the vibrational
state to the MECS 
\begin{multline}
\vert\alpha e^{i\tau_1}\rangle_1\otimes |\alpha e^{i\tau_2}\rangle
_2\otimes \cdots \otimes \vert\alpha e^{i\tau _N}\rangle_N \\
\pm \vert\alpha e^{-i\tau _1}\rangle_1\otimes |\alpha e^{-i\tau _2}\rangle
_2\otimes \cdots \otimes \vert\alpha e^{-i\tau _N}\rangle_N
\end{multline}
up to a normalization constant, where $\tau _k\equiv\tau (-1)^{i_k}$.

Specifically, for all $i_k=0$ and $\tau =\pi /2$, the above state reduces to
the MECS~\cite{Ans94} 
\begin{multline}
\left[ 2\pm 2e^{-2N\vert\alpha |^2}\right]^{-1/2}
	\Big(\vert i\alpha \rangle_1
	\otimes \vert i\alpha \rangle_2\otimes \cdots
	\otimes |i\alpha \rangle_N \\
\pm |-i\alpha \rangle_1\otimes |-i\alpha \rangle_2\otimes \cdots \otimes
\vert -i\alpha \rangle_N\Big)~.
\end{multline}
Therefore, by measuring the combined electronic states of the ions via a
generalized Bell state measurement, the
outcome of the electronic state measurements is a MECS. Preparing the MECS
follows a natural generalization to the entanglement swapping method for
preparing the bipartite ECS.

In order to generate the MECS, we need to perform a (challenging) measurement on certain maximally
entanged electronic states.  Here we provide a
way to realize the measurement using Controlled--NOT (denoted by ${\bf C\!N}$)
gates~\cite{Gates}. A series of ${\bf C\!N}$ gates is applied to the
electronic state~(\ref{eq:e}) followed by a Hadamard ($\bf H$)
gate,
\begin{equation}
{\bf G}={\bf H}_1{\bf C\!N}_{1N}{\bf\ldots C\!N}_{13}{\bf C\!N}_{12},
\end{equation}
where the subscripts of~${\bf C\!N_{ij}}$ denotes the control ion~$i$
and the target~$j$.
Let the state (%
\ref{eq:e}) be the input of this gate ${\bf G}$.
The output is the product
state $\vert j_1\rangle \otimes |j_2\rangle \otimes \cdots \otimes |j_N\rangle
$ $(j_k\in \{0,1\})$.
Local measurements on this output product state after the gate~$\bf G$ has
been applied correspond to the desired generalized Bell state measurements.

One problem that can arise in creating multimode entanglement is that,
during the interaction time between ions,
the state of the vibrational mode may change and no longer exist as
coherent state.  However, thereby exists a remedy that uses the interaction
Hamiltonian for the $i^{\rm th}$ ion with the CoM mode~\cite{Mon96}:
\begin{equation}
{\cal H}_i=-i(\alpha_i a_1^\dagger-\alpha_i^*a_1)(1-\sigma _{iz})
\label{eq:h1}
\end{equation}
during C\!N gate operations, and avoids deleterious changes to the
vibrational mode state~\cite{Mil98}. 
The evolution operator $\exp(-it{\cal H}_i)$ incorporates
the unitary operators
$\exp[\pm ik_xX_1(1-\sigma_{iz})/2]$ and
$\exp[\pm ik_pP_1(1-\sigma_{jz})/2]$ $(i\neq j)$, where
$X_1=(a_1+a_1^\dagger)\sqrt{2}$,
$P_1=(a_1-a_1^\dagger)/\sqrt{2}i$ and $k_x,k_y$ are real numbers.
From these two unitary operators, 
$e^{\pm ik_pP_1(1-\sigma _{jx})/2}$ can be realized by a single--qubit rotation. 
As $[1-\sigma_{iz},1-\sigma_{jx}]=0$,
we can use the technique in Ref.~\cite{Mil98} to realize
\begin{multline}
\exp \left[ -ik_xk_p(1-\sigma_{iz})(1-\sigma _{j,x})/4\right]  \\
=\exp (ik_xX_1(1-\sigma _{iz})/2)\exp (ik_pP_1(1-\sigma _{j,x})/2)  \\
\times\exp (-ik_xX_1(1-\sigma_{iz})/2)	\\
\times\exp (-ik_pP_1(1-\sigma_{jx})/2)
\label{eq:top}
\end{multline}
by an appropriate choice of Raman laser pulse phases.
Letting $k_xk_p=\pi ,$ we obtain
\begin{equation}
{\bf C\!N}_{ij}=\exp \left[ -i{\frac \pi 4}(1-\sigma _{iz})(1-\sigma
_{jx})\right] .
\end{equation}
The vibrational degree of trapped ions acts as a databus, and it does not
change after the gate operation; i.e.,  the
vibrational modes remain as coherent states.

\section{Quantifying the Entanglement}
The method for preparing the MECS has been discussed;
now we quantify the entanglement for this state.
Multipartite entanglement continues to
be an important topic, and here we choose to study entanglement by
certain well--accepted measures, which suffice for the MECS. 

Whereas we have shown how to generate the special case of
even and odd MECSs,
in this section we consider the generalized balanced MECS for studying
entanglement.
By `balanced', we refer to the constraint that each element of the
superposition of multipartite coherent states in the MECS is equally
weighted with all other elements of the superposition.  An unbalanced
ECS is more difficult to construct~\cite{Wie93}, and the extension of
the following work to the unbalanced MECS is straightforward; because
of the greater challenge in creating unbalanced MECSs and the 
ready generalization of the analysis below to unbalanced MECSs, we 
do not include this analysis of unbalanced MECS here.

Thus far, we have considered the even and odd balanced MECS as
generated by a unitary evolution in a larger Hilbert space followed by
measurements on the other degrees of freedom.  A unitary evolution can
be used to generate the bipartite ECS, on the other
hand~\cite{San92,San00}, which does not yield the even and odd variety
of ECS.  The $N$--partite MECS with an arbitrary relative phase~$\theta$ is
given by
\begin{multline}
\vert\alpha ,\theta ,N\rangle_{\rm ECS}
={\cal N}(\vert\alpha \rangle_1\otimes
\vert\alpha \rangle_2\otimes \cdots\otimes \vert\alpha \rangle_N \\
+e^{i\theta }|-\alpha \rangle_1\otimes |-\alpha \rangle_2\otimes
\cdots\otimes |-\alpha \rangle_N)  \label{eq:main}
\end{multline}
with 
\begin{equation}
{\cal N} \equiv \left[2+2p^N\cos\theta\right]^{-1/2}
\end{equation}
the normalization constant and
\begin{equation}
p \equiv e^{-2\vert\alpha\vert^2} = \langle -\alpha \vert \alpha \rangle~.
\end{equation}
The MECS satisfies the equation
\begin{equation}
a_1a_2 \ldots a_N\vert\alpha ,\theta ,N\rangle_{{\rm ECS}}=\alpha ^N|\alpha ,\theta
,N\rangle_{{\rm ECS}},
\end{equation}
where $a_i(i=1,2, \ldots ,N)$ is the annihilation operator of mode $i$.
Two interesting limits arise for $\vert\alpha\vert\rightarrow\infty$
and $\vert\alpha\vert\rightarrow 0$.
In the asymptotic limit $\vert\alpha |\rightarrow \infty $, the two states $|\alpha
\rangle $ and $|-\alpha \rangle $ approach orthogonality, and the state $%
\vert\alpha,\theta,N \rangle_{{\rm ECS}}$ approaches the multipartite maximally entangled
state,
\begin{align}
\vert\text{GHZ}\rangle_N &= \frac 1{\sqrt{2}}(\vert {\bf 0}\rangle_1\otimes |{\bf 0}\rangle_2\otimes
		\cdots \otimes\vert {\bf 0}\rangle_N \nonumber \\
	&\quad +e^{i\theta }\vert {\bf 1}\rangle_1\otimes 
	\vert {\bf 1}\rangle_2\otimes \cdots \otimes
\vert {\bf 1}\rangle_N).
\end{align}
An orthogonal basis can be constructed such that
$\vert {\bf 0}\rangle_i\equiv \vert\alpha\rightarrow\infty\rangle_i$
and $\vert{\bf 1}\rangle_i\equiv
\vert\alpha\rightarrow -\infty \rangle_i,$ where we have symbolically identified the large~$\vert\alpha\vert$ limit.
In the asymptotic limit $|\alpha |\rightarrow
0,$ the state $\vert\alpha ,\pi ,N\rangle_{{\rm ECS}}$ reduces to the so-called 
${\rm W}$ state~\cite{Wan01,Dur00}
\begin{align}
\vert\text{\rm W}\rangle_N   
	&= N^{-1/2}(\vert 1\rangle\otimes_1\vert 0 \rangle_2\otimes \cdots\otimes 
       \vert0\rangle_N\nonumber\\
	&\quad +\vert 0\rangle_1\otimes\vert 1\rangle_2\otimes\ldots\otimes \vert0\rangle 
      +\cdots\nonumber \\
	&\quad + \vert 0\rangle_1\otimes\vert 0\rangle_2 \otimes \cdots\otimes \vert 1\rangle_N )~.
\label{Wstate}
\end{align}
Here $\vert n\rangle_i $ $(n=0,1)$ denote the Fock states of mode $i$.

We employ concurrence~\cite{Hil97} as a measure of bipartite
entanglement for the state $\vert\alpha ,\theta ,N\rangle_{{\rm ECS}}$.
For $\rho_{12}$ the density matrix for a pair of qubits~$1$
and~$2$, the concurrence is~\cite{Hil97} 
\begin{equation}
C_{12}=\max \left\{ \lambda _1-\lambda _2-\lambda _3-\lambda _4,0\right\} 
\label{eq:c1}
\end{equation}
for~$\lambda_1\ge\lambda_2\ge\lambda_3\ge\lambda_4$
the square roots of the eigenvalues of the operator 
\begin{equation}
\varrho_{12}\equiv\rho_{12}(\sigma_y\otimes\sigma _y)\rho_{12}^*(\sigma_y\otimes \sigma_y) \label{eq:c2}
\end{equation}
with $\sigma_y=\left(\begin{smallmatrix}0&-i\\i&0\end{smallmatrix}\right)$
a Pauli matrix.

Nonzero concurrence occurs if and only if qubits 1 and 2
are entangled. Moreover, $C_{12}=0$ only for an unentangled
state, and $C_{12}=1$ only for a maximally entangled state.
This concurrence measure can be extended to multipartite systems,
and we apply it to pure--state and mixed--state entanglement for MECS below.

\subsection{Pure--state entanglement}

In this subsection we consider bipartite splitting of the
multipartite system, i.e., splitting the
entire system into two subsystems, one subsystem containing any~$k$
$(1\le k\le N-1)$ particles and
the other containing the remaining $N-k$ particles.
Let ${\cal C}_{(k,N-k)}$ denote the concurrence between the two subsystems.
Applying the general result for concurrence of bipartite
nonorthogonal states~\cite{Wan01} to the MECS (\ref{eq:main}) yields
\begin{equation}
{\cal C}_{(k,N-k)}(\theta )
	=\frac{\sqrt{(1-p^{2k})(1-p^{2(N-k)})}}{1+p^N\cos\theta}
\label{eq:pure}
\end{equation}
from which the condition for the maximally entangled ECS,
(i.e., ${\cal C}_{(k,N-k)}(\theta )=1$) is given by
\begin{equation}
\cos \theta =-\frac{1-\sqrt{(1-p^{2k})(1-p^{2(N-k)})}}{p^N}~.
\label{eq:cos}
\end{equation}
Hence,
\begin{equation}
1-\sqrt{(1-p^{2k})(1-p^{2(N-k)})}\geq p^N~,  \label{eq:ie}
\end{equation}
with equality only for $2k=N$. 
It follows that the condition for ${\cal C}%
_{(k,N-k)}(\theta )=1$ is simply
\begin{equation}
\cos \theta =-1, \; 2k=N~.
\end{equation}
Thus, for even number~$N$, $k=N/2$, and $\theta=\pi$,
the state $\vert\alpha ,\theta ,N\rangle_{\rm ECS}$ is maximally entangled in the
sense that ${\cal C}_{(N/2,N/2)}(\pi )=1$. For other nonorthogonal 
($p\neq 0$) cases, the state is not maximally entangled.

\subsection{Mixed--state entanglement}

Now we study the bipartite reduced density matrix $\rho_{kl}$, which is
obtained by tracing out all other systems except systems $k$ and $l.\,$There
are $N(N-1)/2$ different density matrices $\rho_{kl}$.
However, for our state 
$\vert\alpha ,\theta ,N\rangle_{\rm ECS}$, all particles are equally entangled with
each other and all the reduced density matrices $\rho_{kl}$ are identical.
Therefore, it is sufficient to consider $\rho_{12}$ and to generalize
from this case.
For convenience, we first make a local transformation
$(-1)^{a_2^{\dagger }a_2}$ on the state $\vert\alpha ,\theta ,N\rangle_{\rm ECS}$.
This local
transformation does not change the amount of entanglement in the state.
Then, by tracing out systems $3,4,\ldots,N$ in the transformed state,
we obtain the reduced density matrix describing systems 1 and 2 as 
\begin{eqnarray}
\rho_{12} &=&\text{Tr}_{3,4,\ldots,N}(\vert\alpha,\theta,N\rangle_{\rm ECS}\langle
\alpha ,\theta ,N|)  \nonumber \\
&=&{\cal N}^2(\vert\alpha \rangle |-\alpha \rangle \langle \alpha |\langle
-\alpha\vert
+\vert -\alpha \rangle \vert\alpha \rangle \langle -\alpha |\langle \alpha |) 
\nonumber \\
&&+e^{i\theta }q|-\alpha \rangle \vert\alpha \rangle \langle \alpha |\langle
-\alpha\vert\nonumber\\
&&+e^{-i\theta }q\vert\alpha \rangle |-\alpha \rangle \langle -\alpha |\langle
\alpha\vert )  \label{eq:mix}
\end{eqnarray}
with $q \equiv p^{N-2}$.

In order to diagonalize the density matrix,
we choose an orthogonal basis $\{\vert {\bf 0}\rangle ,\vert {\bf 1}\rangle \}$,
distinguished from the electronic state basis of the same notation
employed earlier for the entanglement swapping operation by using boldface symbols.
This orthogonal basis is defined as
\begin{equation}
\vert{\bf 0}\rangle\equiv\vert\alpha \rangle~,\;
\vert{\bf 1}\rangle\equiv(\vert -\alpha \rangle -p\vert {\bf 0}\rangle)/{\cal M}~,
\label{eq:alpha}
\end{equation}
where ${\cal M}=\sqrt{1-p^2}$.
It then follows that
\begin{equation}
\vert -\alpha \rangle={\cal M}\vert{\bf 1}\rangle+p\vert{\bf 0}\rangle~.  \label{eq:malpha}
\end{equation}

Substituting Eqs.~(\ref{eq:alpha})-(\ref{eq:malpha})
into Eq.~(\ref{eq:mix}),
we obtain the density matrix
\begin{equation}
\rho_{12} =	{\cal N}^2
\left( \begin{smallmatrix}
2p^2(1+q\cos \theta ) & p{\cal M}(1+qe^{i\theta })
	& p{\cal M} (1+qe^{-i\theta }) & 0 \\ 
p{\cal M}(1+qe^{-i\theta}) & {\cal M}^2 & {\cal M}^2qe^{-i\theta } & 0 \\ 
p{\cal M}(1+qe^{i\theta}) & {\cal M}^2qe^{i\theta } & {\cal M}^2 & 0 \\ 
0 & 0 & 0 & 0
\end{smallmatrix}
\right)
\label{eq:11}
\end{equation}
in the standard basis
$\{\vert{\bf 00}\rangle ,\vert{\bf 01}\rangle ,\vert{\bf 10}\rangle ,
	\vert{\bf 11}\rangle \}$.
From Eqs.~(\ref{eq:c1}) and (\ref{eq:11}), the square roots of eigenvalues
of $\varrho_{12}$ in Eq.~(\ref{eq:c2}) are
\begin{align}
\lambda_1 &={\cal N}^2{\cal M}^2(1+q),\;
\lambda_2 ={\cal N}^2{\cal M}^2(1-q),  \nonumber \\
\lambda_3 &=\lambda_4=0.  \label{eq:eigen}
\end{align}
Although $\rho _{12}$ is complicated,
the expressions for the square roots of the eigenvalues are rather simple.
The concurrence is thus not complicated and follows
directly from Eqs.(\ref{eq:c1}) and (\ref{eq:eigen}), namely
\begin{equation}
{\cal C}_{12}=\frac{{\cal M}^2q}{1+p^N\cos\theta}
	=\frac{p^{N-2}-p^N}
	{1+p^N\cos\theta}~.
\label{eq:mixmix}
\end{equation}

This expression for concurrence must be calculated carefully
for~$p=1$ and $\theta=\pi$.
Of course, $p=1$ implies that $\alpha=0$, which is the limit that
MECS approaches the vacuum state.  In this limit, we apply l'H\^{o}pital's
rule to Eq.~(\ref{eq:mixmix}) to obtain
$\lim_{p\rightarrow 1}{\cal C}_{12} = 2/N$
for~$N>2$.
This result is in accordance with the known maximal degree of entanglement
between any pair of qubits in an $N$--qubit system,
attained for qubits prepared in the pure symmetric state
referred to as the {\rm W} state~\cite{Dur00,Koa00} and presented for
MECS in Eq.~(\ref{Wstate}).
The limit $\vert\alpha\vert\rightarrow 0$ yielding a nonzero
concurrence can thus be understood in the context of producing
a symmetric state~\cite{Wan01}.

The first application of this formula for concurrence is to determine
when systems~1 and~2 are disentangled, i.e., ${\cal C}_{12}=0$.
One case arises for~$p=0$, which corresponds to the orthogonal case.
As described earlier, this case is only valid in the asymptotic limit
of infinite~$\vert\alpha\vert$.
Another case of complete disentanglement arises for $N \rightarrow \infty$ and 
$0<p<1$, yielding a concurrence of ${\cal C}_{12}=0$.
The third case arises for~$\theta \neq \pi$ and in the limit $\vert\alpha|\rightarrow 0$.
In summary, there is no bipartite entanglement in three cases:
the asmptotic limit of infinite--amplitude coherent states, the
asymptotic limit of an infinite number of entangled systems, and
the case of the MECS for which the coherent state is just the vacuum state.

For $N=2$ the concurrence (\ref{eq:mixmix}) reduces to
\begin{equation}
{\cal C}_{12}=\frac{1-p^2}{1+p^2\cos\theta},
\label{eq:m}
\end{equation}
which is the concurrence for the pure state $\vert\alpha, \theta, 2\rangle_{\rm ECS}$.
The bipartite concurrence for a bipartite ECS provides arbitrarily strong
entanglement for appropriate parameter choices.  When $\theta=\pi$, the concurrence
${\cal C}_{12}=1$, and the state becomes the the antisymmetric state $|\Psi^-\rangle$ 
(\ref{eq:bell}). Bipartite entanglement
of multipartite systems offers reduced entanglement, however.
For $N \neq 2$, the reduced density matrix describes a mixed state, and the
degree of entanglement is given by Eq.~(\ref{eq:mixmix}).

Figure 1 gives a plot of the concurrence verus $\theta$ and $p$.
As seen from the figure, the maximum value of the concurrence occurs
when $\theta =\pi $ for fixed values of $p$ and $N$. From Eq.~(\ref{eq:mixmix}), the maximum value
is obtained as
\begin{equation}
{\cal C}_{12}=\frac{p^{N-2}-p^N}{1-p^N}~.
\end{equation}
\begin{figure}
\includegraphics*[width=3in,keepaspectratio]{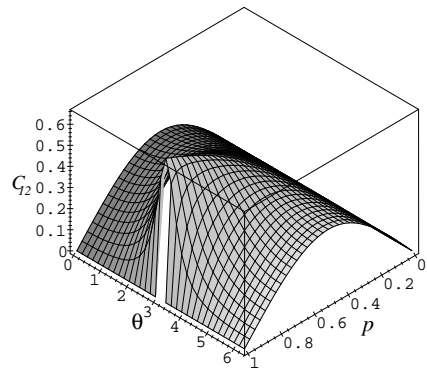}
%\epsfig{width=8cm,file=eswapfig1.ps}
%\\
\caption[]{The concurrence versus $\theta$ and $p$ for $N=3$.} 
\end{figure}

As a short summary we give concurrences of some special states in Table I
\begin{table}
\caption{A summary of concurrences for some special states}
\begin{tabular}{cccccc}
$\vert\alpha|$&  $p$       & $ N $  & $\theta$ & $ C $  & State    \\ \hline
$0$       &  $1$       &   $>2$    & $\pi$ &  $2/N$       &  $|{\rm W}\rangle_N$      \\ 
$0$       &  $1$       &   2    & $\pi$ &      $1$        &   $|\Psi^-\rangle$     \\ 
$\infty$  &  $0$       &   2    & any &  $1$            &  $|\Phi\rangle$  \\ 
$\infty$  &  $0$       &   $>2$ & any &  $0$             &$|{\rm GHZ}\rangle_N$  \\  
$0$       &  $1$       &   $\ge 2$    & $\neq \pi$ &  0        & $\vert 0\rangle^{\otimes N}$\\
$\neq 0,\infty$         &  $0<p<1$         &   $\infty$    & any & 0 &$\vert\alpha,\theta,\infty\rangle_{\rm ECS} $     
\end{tabular}
\end{table}
In Table~I the state $|\Phi\rangle=2^{-1/2}(\vert{\bf 0}\rangle
	\otimes \vert{\bf 0}\rangle
	+\exp(i\theta)\vert{\bf 1}\rangle\otimes\vert{\bf 1}\rangle )$,
where $|{\bf 0}\rangle\equiv \vert\alpha\rightarrow\infty \rangle$ and $|{\bf 1}
\rangle\equiv |\alpha \rightarrow -\infty \rangle.$

For fixed $N$ and $\theta$ there still exists maximum values of the concurrence
(see Fig.~1). From Eq.~(\ref{eq:mixmix}),
the value of $p$ at which the maximum  occurs is determined by 
the equation
\begin{equation}
2p^N\cos\theta+Np^2=N-2.
\end{equation}
As an example, we consider the tripartite case $N=3$.
The above equation simplifies to
\begin{equation}
2p^3\cos \theta +3p^2-1=0.  \label{eq:three}
\end{equation}
For $\theta =0$ and $0<p<1$,
the solution is $p=1/2$ with a corresponding maximum concurrence
of $1/3$. For $\theta=\pi/2$, the solution of for~$p$ is $3^{-1/2}$,
and the maximum value of ${\cal C}_{12}$ is $2\sqrt{3}/9$.
%\footnote{}

\subsection{Multipartite entanglement}

We have thus far considered only bipartite entanglement of a multipartite
system. One type of multipartite entanglement is $N$--way entanglement which
involves all $N$ particles. We have used the concurrence to example bipartite
entanglement. Recently Coffman {\it et al}~\cite
{Cof00} used concurrence to examine three-qubit systems, and introduced the
concept of the 3--tangle, $\tau _{1,2,3}(\vert\psi \rangle )$ 
as a way to
quantify the amount of 3--way entanglment in three-qubit systems. Later Wong
and Christensen~\cite{Won01} generalize 3--tangle to $N$--tangle. The $N$%
--tangle is the square of the multiqubit concurrence 
\begin{equation}
C_{1,2,\ldots ,N}\equiv |\langle \psi |\sigma _y^{\otimes N}\vert\psi
^{*}\rangle |  \label{eq:m1}
\end{equation}
for even qubits, with~$\vert\psi \rangle $ a multiqubit pure state. This
concurrence works only for even numbers of qubits;
$\vert\langle \psi |\sigma _y^{\otimes N}\vert\psi
^{*}\rangle\vert=0$ for any odd-$N$ qubit pure states.
Therefore, this quantity cannot
act as a general measure of $N$--way entanglement. Next we quantify the $N$--way
entanglment using $N$--tangle for $N=3$ and even $N.$ 

The 3--tangle can be calculated from concurrences $C_{1(23)},C_{12},$ and $%
C_{13}$ because~\cite{Cof00}
\begin{equation}
\tau _{1,2,3}=C_{1(23)}^2-C_{12}^2-C_{13}^2  \label{eq:tau}
\end{equation}
holds. For our state $\vert\alpha ,\theta ,3\rangle_{{\rm ECS}}$,
the 3--tangle is
simplified as
\begin{equation}
\tau _{1,2,3}=C_{1(23)}^2-2C_{12}^2.  \label{eq:tau2}
\end{equation}
From Eqs.~(\ref{eq:pure}) and (\ref{eq:mixmix}), the 3--tangle is
\begin{equation}
\tau _{1,2,3}=\frac{(1-p^2)^3}{(1+p^3\cos \theta )^2}~,
\label{eq:tau123}
\end{equation}
and, as expected,
$\tau _{1,2,3}=1$ ($\tau _{1,2,3}=0$) in the limit 
that $p\rightarrow 0$($p\rightarrow 1$).

Now we examine $N$--way entanglement for the state $\vert\alpha ,\theta
,N\rangle_{{\rm ECS}}$ with $N$ even. In the basis of Eq.~(\ref{eq:alpha}), the state $\vert\alpha ,\theta ,N\rangle_{{\rm ECS}}$ can be
rewritten as 
\begin{multline}
\vert\alpha ,\theta ,N\rangle_{{\rm ECS}}
	= {\cal N}(\vert{\bf 0}\rangle_1\otimes
	\vert{\bf 0}\rangle_2\otimes\cdots
	\otimes\vert{\bf 0}\rangle_N \\
	+e^{i\theta }({\cal M}\vert{\bf 1}\rangle_1
	+p\vert{\bf 0}\rangle_1)\otimes ({\cal M}
	\vert{\bf 1}\rangle_2+p\vert{\bf 0}\rangle_2) \\
	\otimes \cdots \otimes ({\cal M}\vert{\bf 1}\rangle_N
	+p\vert{\bf 0}\rangle_N))~.
\label{eq:m2}
\end{multline}
The effect of writing the MECS in this basis set is to have the state
expressed formally as a multiqubit state. From Eqs.~(\ref{eq:m1}) and (\ref
{eq:m2}) the $N$--tangle is obtained as 
\begin{equation}
\tau _{1,2,\ldots ,N}=\frac{(1-p^2)^N}{(1+p^N\cos \theta )^2},
\label{eq:ccc}
\end{equation}
for even $N$. Although this formula is obtained for even $N,$ by comparing
Eqs.~(\ref{eq:tau123}) and (\ref{eq:ccc}), it is also applicable to $N=3.$  
The condition for maximal entanglement, $\tau _{1,2,\ldots ,N}=1$, is 
\begin{equation}
N=2,\;\cos \theta =-1,
\end{equation}
for $p\neq 0$. This constraint on~$N$ restricts maximal entanglement to the
bipartite ECS.

In Fig.~2 we give a plot of the $N$--tangle vs~$p$
for various $\theta$ and
$N$. For $p=0$ (orthogonal case), the multiqubit concurrence is
equal to 1 independent of $\theta $. We already know that the
state $\vert\alpha
,\pi ,N\rangle_{{\rm ECS}}$ becomes the~$W$ state in the limit $p\rightarrow 1$%
. Now we take this limit and choose $\theta =\pi $ in Eq.~(\ref{eq:ccc}),
thereby establishing that the concurrence $\tau _{1,2,\ldots ,N}=0$ in this
case. Thus, we observe that multipartite entanglement,
as determined by the $N$--tangle, is indeed zero for the $W$ state.

\begin{figure}
\includegraphics*[width=3in,keepaspectratio]{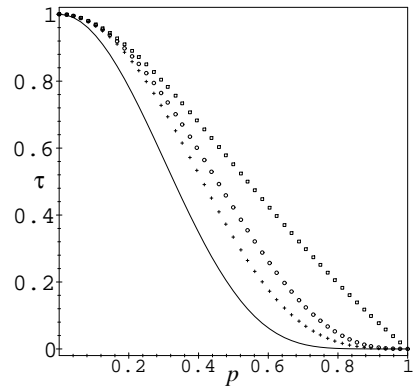}
%\epsfig{width=8cm,file=eswapfig2.ps}
%\\
\caption[]{The $N$--tangle versus $p$ for different $\theta$ and $N$: $N=3, \theta=0$ (cross points), 
$N=3, \theta=\pi/2$ (circle points),$N=3, \theta=\pi$ (box points), and 
$N=6, \theta=0$ (solid line).} 
\end{figure}

\section{Conclusion}

We have considered generation and entanglement measures for
the multipartite entangled coherent state. 
Generating the MECS is possible by entangling vibrational degrees
of freedom for trapped two--level ions with the ions' internal electronic
states. By measuring the electronic states with respect to a highly
entangled basis, basically an extension of the (bipartite) Bell basis,
the resultant motional state is a MECS.
We have quantified the entanglement of the MECS by applying the concurrence
to measure bipartite entanglement (in one case by splitting the multipartite
system into two subsystems and, in the other case, by tracing over all 
degrees of freedom except for two subsystems).  We have also employed the
$N$--tangle to determine the overall degree of entanglement. Each of
these measures tells us something important about the MECS,
and the MECS versions of the GHZ and W states have been studied and
elucidated.

Quantifying entanglement for MECSs provides a simple 
measure to evaluate the inherent resource of such states, and this is
relevant to quantum information applications where entanglement is
regarded as a crucial resource.  Moreover, the study of MECS highlights the
subtleties of applying entanglement measures to nonorthogonal entangled
states.  The particular physical realization
studied here has been entangled vibrational motion of ions in a trap.
As the bipartite ECS has proven to be a useful alternative construct
for making qubits~\cite{Mun00},
in contrast to the usual Fock state qubits, this analysis
could be valuable for encoding qubits as MECS in an ion trap.

Finally, the analysis here for MECS is readily exteneded to more general
systems, including entangled squeezed states~\cite{San95},
entangled SU(2) and SU(1,1) coherent states~\cite{Wan00} and so on,
as follows.
Essentially, Eqs.~(\ref{eq:pure}), (\ref{eq:mixmix}) and (\ref{eq:ccc})
can be applied to the general entangled state
\begin{multline}
\vert\Psi ,\Phi \rangle ={\cal N}^{\prime }(\vert\Psi\rangle_1\otimes\vert\Psi
\rangle_2\otimes \cdots\otimes\vert\Psi \rangle_N  \\
+e^{i\theta ^{\prime }}|\Phi \rangle_1\otimes |\Phi \rangle_2\otimes
\cdots\otimes\vert\Phi\rangle_N),
\label{eq:last}
\end{multline}
for ${\cal N}^{\prime }$ the normalization constant, $|\Psi \rangle $
and $|\Phi \rangle $ arbitrary linearly independent states, and
$\langle \Psi |\Phi \rangle =p^{\prime }$ a real overlap.  Then the
corresponding concurrence for the state $|\Psi ,\Phi \rangle $ is obtained
by directly replacing $p$ and $\theta $ with $p^{\prime }$ and $\theta
^{\prime }$ in Eqs.(\ref{eq:pure}), (\ref{eq:mixmix}) and (\ref{eq:ccc}).
Therefore, our results for quantifying entanglement provide a useful
formalism with a validity well beyond that for MECS.

\acknowledgments
XW appreciates helpful discussions with K.\ M\o lmer and A.\
S\o rensen. The authors also gratefully acknowledge
valuable comments by the anonymous referee.
This work has been supported by the Information Society Technologies
Programme IST-1999-11053, EQUIP action line 6-2-1, the Australian Research Council, and Project Q-ACTA.


\begin{thebibliography}{99}
\bibitem {Ben93}  C.\ H.\ Bennett, G.\ Brassard, C.\ Cr\'{e}peau, R.\ Jozsa,
A.\ Peres, and W.\ K.\ Wootters, \prl {\bf 70}, 1895 (1993).
\bibitem {Ben92}  C.\ H.\ Bennett and S.\ J.\ Wiesner, \prl {\bf 69}, 2881
(1992).
\bibitem {Eke91}  A.\ K.\ Ekert, \prl {\bf 67}, 661 (1991).
\bibitem {Mur99}  M.\ Murao, D.\ Jonathan, M.\ B.\ Plenio, and V.\ Vedral, %
\pra {\bf 59}, 156 (1999).
\bibitem {Wer89}  R.\ F.\ Werner, \pra {\bf 40}, 4277 (1989); S.\ Popescu, %
\prl {\bf 72}, 797 (1994); A.\ Peres, \prl {\bf 77}, 1413 (1996); P.\
Horodecki, Phys.\ Lett.\ A {\bf 232}, 333 (1997); W.\ D\"{u}r, J.\ I.\
Cirac, and R.\ Tarrach, \prl {\bf 83}, 3562 (1999); D.\ Bru\ss , \pra {\bf 60%
}, 4344 (1999); L.\ M.\ Duan, G.\ Giedke, J.\ I.\ Cirac, and P.\ Zoller, 
\prl 
{\bf 84}, 2722 (2000); P.\ Zanardi, \pra {\bf 63}, 040204 (2001); P.\ W.\
Shor, J.\ A.\ Smolin, and B.\ M.\ Terhal, \prl {\bf 86}, 2681 (2001); M.\
Lewenstein, B.\ Kraus, P.\ Horodecki, and J.\ I.\ Cirac, \pra
{\bf 63}, 044304 (2001).
\bibitem {Hil97}  S.\ Hill and W.\ K.\ Wootters, \prl {\bf 78}, 5022 (1997);
W.\ K.\ Wootters, \prl {\bf 80}, 2245 (1998).
\bibitem {Fuc97}  C. A. Fuchs, \prl {\bf 79}, 1162 (1997).
\bibitem {San92}  B.\ C.\ Sanders, \pra {\bf 45}, 6811 (1992);
	{\bf 46}, 2966 (1992).
\bibitem {Wie93}  B.\ Wielinga and B.\ C.\ Sanders, J.\ Mod.\ Opt.\ {\bf 40},
1923 (1993).
\bibitem {Man95}  A.\ Mann, B.\ C.\ Sanders, and W.\ J.\ Munro,
	\pra {\bf 51}, 989 (1995).
\bibitem {San95}  B.\ C.\ Sanders, K.\ S.\ Lee, and M.\ S.\ Kim, \pra
{\bf 52}, 735 (1995).
\bibitem {San00}  B.\ C.\ Sanders and D.\ A.\ Rice, \pra {\bf 61}, 013805
(2000).
\bibitem {Cha92}  C.\ L.\ Chai, \pra {\bf 46}, 7187 (1992).
\bibitem {Ans94}  N.\ A.\ Ansari and V.\ I.\ Man'ko, \pra {\bf 50}, 194
(1994); V.\ V.\ Dodonov, V.\ I.\ Man'ko, and D.\ E.\ Nikonov, \pra {\bf 51},
3328 (1995); S.\ B.\ Zheng, Quantum Semiclass.\ Opt.\ {\bf 10}, 691 (1998).
\bibitem {Wan00}  X.\ Wang, B.\ C.\ Sanders, and S.\ H. Pan, J.\ Phys.\ A:
Math.\ Gen.\ {\bf 33}, 7451 (2000).
\bibitem {Mun00}  W.\ J.\ Munro, G.\ J.\ Milburn, and B.\ C.\ Sanders, \pra 
{\bf 62}, 052108 (2000).
\bibitem {Cof00}V.\ Coffman, J.\ Kundu, and 
W.\ K.\ Wootters, \pra {\bf 61}, 052306 (2000).
\bibitem {Won01}  A.\ Wong and N.\ Christensen, \pra {\bf 63}, 044301 (2001).
\bibitem {Jam98} D.~F.~V.~James, App.\ Phys.\ B: Lasers and Optics
	{\bf 66}, 181 (1998).
\bibitem {Kie01} D.~Kielpinski, A.\ Ben-Kish, J.\ Britton, V.\ Meyer,
	M.\ A.\ Rowe, C.\ A.\ Sackett, W.\ M.\ Itano, C.\ Monroe,
	and D.\ J.\ Wineland, quant-ph/0102086.
\bibitem {Mon96}  C.\ Monroe, D.\ M.\ Meekhof, B.\ E.\ King, and D.\ J.\
Wineland, Science {\bf 272}, 1131 (1996).
\bibitem {Ger97}  C.\ C.\ Gerry, \pra {\bf 55}, 2487 (1997).
%\bibitem {Sav90}  C.\ M.\ Savage, S.\ L.\ Braunstein, and D.\ F.\ Walls, Opt.
%Lett.\ {\bf 15}, 628 (1990).
%\bibitem {Bru92}  M.\ Brune, S.\ Haroche, J.\ M.\ Raimond, L.\ Davidovich and
%N.\ Zagury, \pra {\bf 45}, 5193 (1992); M.\ Brune, E.\ Hagley, J.\ Dreyer,
%X.\ Ma\^{i}tre, A.\ Maali, C.\ Wunderlich, J.\ M.\ Raimond, and S.\ Haroche, %
%\prl {\bf 77}, 4887 (1996).
\bibitem {Zuk93}  M.\ \.{Z}ukowski, A.\ Zeilinger, M.\ A.\ Horne, and A.\
K.\ Ekert, \prl {\bf 71}, 4287 (1993); J.\ W.\ Pan, D.\ Bouwmeester, H.\
Weinfurter, and A.\ Zeilinger, \prl {\bf 80}, 3891 (1998); S.\ Bose, V.\
Vedral, and P.\ L.\ Knight, \pra {\bf 60}, 194 (1999).
\bibitem {Gates}  A. Barenco, C. H. Bennett, R. Cleve, D. P. DiVincenzo, N.
Margolus, P. Shor, T. Sleator, J. A. Smolin, and H. Weinfurter, \pra {\bf 52}%
, 3457 (1995); J. I. Cirac and P. Zoller, \prl {\bf 74}, 4091 (1995).
\bibitem {Mil98}  G. J. Milburn, quant-ph/9908037; A. S\o rensen and K. M\o %
lmer, \pra {\bf 62}, 022311 (2000); X. Wang, A. S\o rensen and K. M\o lmer, %
\prl {\bf 86}, 3907 (2001);
\bibitem {Wan01}  X.\ Wang, quant-ph/0102011; X.\ Wang, Phys. Rev. A {\bf 64}, 022303 (2001).
\bibitem {Dur00}  W.\ D\"{u}r, G.\ Vidal, and J.\ I.\ Cirac, \pra {\bf 62},
062314 (2000); W.\ D\"{u}r, \pra {\bf 63}, 020303 (2001).
\bibitem {Koa00}  M.\ Koashi, V.\ Bu\v {z}ek, and N.\ Imoto, \pra {\bf 62},
050302 (2000).
\end{thebibliography}
\end{document}